\documentclass[12pt]{article}
\usepackage{latexsym,amsfonts,amsmath,amsthm,amssymb}

\title{Gene expression from polynomial dynamics in the $2$-adic information space}

\author{Andrei Yu. Khrennikov\\
Center for Mathematical
Modeling \\ in Physics and Cognitive Sciences,\\
University of V\"axj\"o, S-35195, Sweden\\
Email:Andrei.Khrennikov@msi.vxu.se}

\begin{document}

\maketitle

\abstract{We perform geometrization of genetics by representing genetic information by
points of the 4-adic {\it information space.} By well known theorem of number
theory this space can also be represented as the 2-adic space. The process of
DNA-reproduction is described by the action of a 4-adic (or equivalently
2-adic) dynamical system. As we know, the genes contain information for
production of proteins. The genetic code is a degenerate map of codons to
proteins. We model this map as functioning of a polynomial dynamical system.
The purely mathematical problem under consideration is to find a dynamical
system reproducing the degenerate structure of the genetic code. We present one
of possible solutions of this problem.}

\section{Introduction}

During last ten years there were found numerous applications of $p$-adic numbers outside 
the domain of number theory -- in particular, in quantum physics, Beltrametti and Cassinelli, 
1972, Volovich, 1987, Vladimirov et al, 1994, Khrennikov 1994,  and 
theory of disordered systems, Avetisov et al, 1999a,b, 2002 a,b, Parisi and Sourlas,  2000,
Kozyrev et al, 2005, 
Khrennikov and Kozyrev, 2006 a,b. We pay attention to the series
of papers, Khrennikov, 1997, 1998 a,b, 1999, 2000 a,b, and 2004 a,b 
in that {\it $p$-adic information space} was introduced and applied to
information theory, cognitive and social sciences,  psychology and neurophysiology, see also 
Pitk\"anen, 1998, Khrennikov and Nilsson, 2004. The main distinguishing feature of encoding 
of information by $p$-adic 
numbers is the possibility to encode the hierarchical structure of information through 
the ultrametric topology on the $p$-adic tree, cf. also with 
Voronkov, 2002 a,b. As was mentioned, this possibility was explored 
a lot in Khrennikov, 1997, 1998 a,b, 1999, 2000 a,b, and 2004 a,b. Recently 
it was pointed out that the same {\it $p$-adic information space} 
can be applied to mathematical modeling of gene expression, Dragovich, 2006, and
Khrennikov 2006 a,b.

We apply this approach to genetics.
Now we present schematically development of this model. DNA and RNA sequences are represented by
4-adic numbers. Nucleotides are mapped to digits in registers of
4-adic numbers: thymine - $T=0,$ cytosine - $C=1,$ adenine - $A=2,$ and guanine - $G=3.$ 
 The $U$-nucleotide is
represented (as well as $T)$ by 0.\footnote{As an introduction to modern 
problems of genetics one can 
use the special issue of Nature Collection, 2006.} 
The {\it DNA and RNA sequences have the natural
hierarchical structure: letters which are located at the beginning of
a chain are considered as more important.} This hierarchical structure
coincides with the hierarchical structure of the 4-adic tree. Such a hierarchy can
also be encoded by the 4-adic metric. The process of
DNA-reproduction is described by the action of a 4-adic dynamical system.
As we know, the genes contain information for production of
proteins. The genetic code is a degenerate map of codons to
proteins. We model this map as functioning of a polynomial 4-adic
dynamical system. Proteins are associated with cycles
of such a dynamical system. By well known theorem of number theory 
this dynamics can also be represented in
the 2-adic space. 

\section{$m$-adic ultrametric spaces}

The notion of a metric space is used in many applications for
describing distances between objects. Let $X$ be a set. A function
$\rho: X\times X \to {\bf R}_+ $ (where ${\bf R}_+ $ is the set of
positive real numbers) is said to be a {\it metric} if it has the
following properties: $ 1) \rho(x,y)= 0 \; \; \mbox{iff} x=y \; \;
\mbox{(non-degenerated)};  2)\rho(x,y)=\rho(y,x)
\;\;\mbox{(symmetric)}; 3) \rho(x,y)\leq \rho(x,z) +\rho(z,y)\; \;
\mbox{(the triangle inequality).}$  The pair $(X,\rho)$ is called a
metric space.

We are interested in the following class of metric spaces
$(X,\rho).$ Every point $x$ has the infinite number of coordinates
\begin{equation}
\label{K} x=(\alpha_1,...,\alpha_n,...)\;.
\end{equation}
Each coordinate yields the finite number of values,
\begin{equation}
\label{K1} \alpha\in A_m=\{ 0,...,m-1\},
\end{equation}
where $m > 1$ is a natural number, the base of the alphabet $A_m.$
The metric $\rho$ should be so called ultrametric, i.e.,
satisfy the {\it strong triangle inequality}:
\begin{equation}
\label{K2} \rho (x,y)\le\max [\rho (x,z),\rho (z,y)], \; x,y,z\in X.
\end{equation}
The strong triangle inequality can be stated geometrically: {\it
each side of a triangle is at most as long as the longest one of the
two other sides.} It is impossible to imagine such a `triangle' in
the ordinary Euclidean space.

We denote the space of sequences (\ref{K}), (\ref{K1}) by the symbol
${\bf Z}_m.$ The standard ultrametric is introduced on this set in
the following way. For two points

$x=(\alpha_0, \alpha_1, \alpha_2,...., \alpha_n,.....),
y=(\beta_0,\beta_1, \beta_2, ..., \beta_n,...)\in {\bf Z}_m,$

we set
$$
\rho_m(x,y)= \frac{1}{m^k} \; \; \mbox{if}\; \;  \alpha_j= \beta_j,
j=0,1,...,k-1,\; \;  \mbox{and} \; \;\alpha_k\not=\beta_k.
$$
This is a metric and even an ultrametric.  To find the distance
$\rho_m(x,y)$ between two strings of digits $x$ and $y$ we have to
find the first position $k$ such that strings have different digits
at this position.

Geometrically we can imagine a system of $m$-adic integers  (which
will be the mathematical basis of our cognitive models) as a
homogeneous {\it tree} with $m$-branches splitting at each vertex.
 The distance between mental states is determined by
the length of their common root: close mental states have a long
common root.  The corresponding geometry strongly differs from the
ordinary Euclidean geometry.
\begin{center}
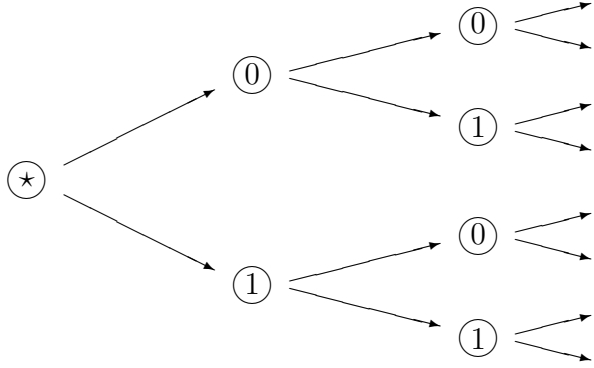
\begin{figure}[ht] \unitlength1cm
\begin{picture}(12,5)
\put(2,3){\circle{0.5}}            \put(1.9,2.9){$\star$}
\put(2.5,2.8){\vector(2,-1){2.0}} \put(2.5,3.2){\vector(2,1){2.0}}
\put(5,4.4){\circle{0.5}}            \put(4.9,4.28){$0$}
\put(5.5,4.35){\vector(4,-1){2.0}} \put(5.5,4.45){\vector(4,1){2.0}}
\put(5,1.6){\circle{0.5}}            \put(4.9,1.48){$1$}
\put(5.5,1.55){\vector(4,-1){2.0}} \put(5.5,1.65){\vector(4,1){2.0}}
\put(8,5.05){\circle{0.5}}          \put(7.9,4.95){$0$}
\put(8,3.7){\circle{0.5}}          \put(7.9,3.57){$1$}
\put(8,2.25){\circle{0.5}}          \put(7.9,2.15){$0$}
\put(8,0.9){\circle{0.5}}          \put(7.9,0.75){$1$}
\put(8.5,5.00){\vector(4,-1){1.0}} \put(8.5,5.10){\vector(4,1){1.0}}
\put(8.5,3.65){\vector(4,-1){1.0}} \put(8.5,3.75){\vector(4,1){1.0}}
\put(8.5,2.20){\vector(4,-1){1.0}} \put(8.5,2.30){\vector(4,1){1.0}}
\put(8.5,0.85){\vector(4,-1){1.0}} \put(8.5,0.95){\vector(4,1){1.0}}
\end{picture}
\caption{The $2$-adic tree}
\end{figure}
\end{center}

Let $(X,\rho)$ be an arbitrary ultrametric space. For  $r\in{\bf
R}_{+},a\in X,$ we set
  $$U_r(a)=\{ x\in X:\rho(x,a)\le r \},\; \;
  U_r^-(a)=\{ x\in X:\rho(x,a) < r\} .
  $$
  These are {\it balls}
  of radius $r$ with center $a.$  Balls have the following
  properties:

1) Let $U$ and $V$ be two balls in $X.$ Then there are only two
possibilities: (a) balls are ordered by inclusion (i.e., $U\subset
V$ or $V\subset U$); (b) balls are disjoint.\footnote{There is the
third possibility in the Euclidean space .}

2) Each point of a ball may serve as a centre.

3) In some ultrametric spaces a ball may have infinitely many radii.

Let $m > 1$ be the fixed natural number. We consider the $m$-adic
metric space $({\bf Z}_m, \rho_m).$  This metric space has the
natural algebraic structure.

A point $x=(\alpha_0, \alpha_1, \alpha_2,..., \alpha_n,....)$ of the
space ${\bf Z}_m$ can identified with a so called $m$-adic number:
\begin{equation}
\label{l1} x= \alpha_0 \alpha_1...\alpha_k....\equiv \alpha_0
+\alpha_1 m +...+ \alpha_k m^k +... \;.
\end{equation}
The series (\ref{l1}) converges in the metric space ${\bf Z}_m.$ In
particular, a finite  string $x= \alpha_0 \alpha_1...\alpha_k$
can be identified with the natural number
$$
x= \alpha_0+  \alpha_1 m+...+\alpha_k m^k.
$$

It is possible to introduce algebraic operations on the set of
$m$-adic numbers ${\bf Z}_m,$ namely addition, subtraction, and
multiplication. These operations are natural extensions by the
$m$-adic continuity of the standard operations on the set of natural
numbers ${\bf N}=\{ 0,1,2,3,...\}.$

\section{Mental information space}

We shall use the following mathematical model for mental information space:

\medskip

(1) Set-structure: The set of mental states $X_{\rm{mental}}$ has
the structure of the $m$-adic tree: $X_{\rm{mental}}= {\bf Z}_m.$

\medskip

(2) Topology: Two mental states $x$ and $y$ are close if they have
sufficiently long common root. This topology is described by the
metric $\rho_m.$

\medskip

In our mathematical model {\it mental space is represented as the
metric space $({\bf Z}_m, \rho_m).$}

\section{Genetic information space}

Genetic information space arises as a special case of mental information spaces.

\subsection{4-adic encoding of nucleotides}

We shall use the following mathematical model for genetic information space.
We choose the 4-adic representation for DNA and RNA:
$$
T=0, C=1, A=2, G=3
$$
and
$$
U=0, C=1, A=2, G=3.
$$
An arbitrary gene in a DNA-sequence is encoded by a 4-adic integer, for example:
$$
ATCGTA... \to 201302...= 2+ 4^2+ 3\cdot 4^3+ 2\cdot 4^5+...
$$ 

Of course, biologically realizable sequences are finite (but very long). Thus they correspond to 
natural numbers. But in a mathematical model we can use even infinitely long genetic 
sequences. Denote this space by the symbol $X_{\rm{genetic}}.$ This space 
has the following distinguishing features:

\medskip

(a) {\it Set-structure:} The set of genetic states $X_{\rm{genetic}}$ has
the structure of the $4$-adic tree: $X_{\rm{genetic}}= {\bf Z}_4.$

\medskip

(b) {\it Topology:} Two genetic states $x$ and $y$ are close if they have
sufficiently long common root.\footnote{Thus the first SNP (single nucleotide polymorphism) 
distinguishes two genetic states.} This topology is described by the
metric $\rho_4.$

\medskip

(c) {\it Dynamics}: Information processing on the level of genetic states is 
described by 4-adic dynamical systems. In the simplest case
of the discrete-time dynamics these are iteration of a map
$$
f: {\bf Z}_4 \to {\bf Z}_4.
$$

(d) {\it Hierarchical structure:} The coding system which is
used our model for recording vectors of information generates a
{\it hierarchical structure} between digits of these vectors -- 
between  nucleotides in the gene-sequence. Thus
if $x=(\alpha_1, \alpha_2,...,\alpha_n,...), \alpha_j=0,1,2,3$
is an information vector which presents genetic information
then digits $\alpha_j$ have different weights. The digit $\alpha_0$
is the most important, $\alpha_1$ dominates over $\alpha_2,...,
\alpha_n,...,$ and so on.

\subsection{Transcription-map}

Transcription is the process of copying a gene into RNA. This is the first step of turning
a gene into protein (although not all transcriptions lead to proteins).
In our coding system transcription is simply the identity map from
${\bf Z}_4 \to {\bf Z}_4$ (since the $T$ and $U$ nucleotides are represented by the same digit).

\section{Genetic code}

\subsection{Encoding of proteins by codons}

In the genetic code proteins are encoded by {\it codons} -- blocks of the length 3 in the 
gene transcription. Each codon contains information for producing of a single amino acid.
By using our 4-adic coding system we can rewrite the table of the genetic code,
see, e.g., Wikipedia, 2006. We collect amino acids in families with respect to a number of codons which are 
used to encode an amino acid:

\medskip

(1). Met: 203; Trp: 033;

\medskip

(2). Asn: 220, 221; Asp: 320, 321;  Cys: 030, 031;  Gln: 122, 123; 

Glu: 322, 323; His: 120, 121;  Lys: 222, 223;  Phe: 000, 001; 

Tyr: 020, 021; 

\medskip

(3) Ile: 200, 201, 202; Stop: 023, 032, 022;

\medskip

(4). Ala: 310, 311, 312, 313;  Gly: 330, 331, 332, 333;

Pro: 110, 111, 112, 113; Thr: 210, 211, 212, 213; 

Val: 300, 301, 302, 303;

\medskip

(6). Arg: 130, 131, 132, 133, 232, 233; 

Leu: 002, 003, 100, 101, 102, 103;

\medskip

Ser: 010, 011, 012, 023, 230, 231;

\subsection{Codon-map} 

First we consider the standard left-shift:
$$
s_l(\alpha_0 \alpha_1 \alpha_2...)=\alpha_1 \alpha_2...
$$
We also consider the following cutoff-map
$$
c_3(\alpha_0 \alpha_1 \alpha_2...)= \alpha_0 \alpha_1 \alpha_2.
$$
Then the representation by codons of the gene-expression is given by 
the $c_3$-projections of the iterations of the left-shift:
$$
x_n=c_3(s_l^{3(n-1)}(x)).
$$

\section{2-adic encoding of proteins}

The 4-adic encoding can be easily transformed into the 2-adic encoding just by using the 2-adic 
representation of the genetic alphabet:

2-adic code: U=00, A=01, C=10, G=11:

We again collect amino acids in families with respect to a number of codons which are 
used to encode an amino acid:

\medskip

(1).  Met: 010011; Trp: 001111;

\medskip

(2). Asn: 010100, 010110; Asp: 110100, 110110; Cys: 001100, 001110; 

Gln: 100101, 100111; Glu: 110101, 110111; His: 100100, 100110;

Lys: 010101, 010111;  Phe: 000000, 000010; Tyr: 000100, 000110;

\medskip

(3).  Ile: 010000, 010010, 010001; Stop: 000111, 001101, 000101; 

\medskip

(4). Ala: 110100, 110101, 110110, 110111; 

Gly: 111100, 111110, 111101, 111111; 

Pro: 101000, 101010, 101001, 101011; 

Thr: 011000, 011010, 011001, 011011; 

Val: 110000, 110010, 110001, 110011;

\medskip

(6).  Arg: 011100, 011101, 011101, 011111, 101110, 101111; 

Leu: 000001, 000011, 100000, 100010, 100001, 100011; 
 
Ser: 001000, 001010, 001001, 000111, 011100, 011110.
 
\section{Dynamical model for degeneracy of the genetic code}

We shall use study dynamical systems corresponding to maps:
\begin{equation}
\label{d1}
{\bf Z}_m \to {\bf Z}_m, x \to f(x).
\end{equation}
As usual, we study the behaviour of iterations $ x_n = f^n(x_0), x_0 \in {\bf Z}_p,$
where $f^n(x)=f \circ ...\circ f(x)= f(....(f(f(x)...),$ 
the result of $n$ successive applications of the map $f.$
We shall use the standard terminology of the theory of dynamical
systems. If $f(x_0)=x_0$ then $x_0$ is a {\it fixed point}.
If $x_n=x_0$  for some $n=1,2,...$ we say that $x_0$ is a {\it periodic
point}. If $n$ is the smallest natural number with this property then $n$
is said to be the {\it period} of $x_0.$ 
We denote the corresponding {\it cycle}
by $\gamma=(x_0, x_1,...,x_{n-1}).$ In particular, the fixed point $x_0$ is the 
periodic point of period 1. Obviously $x_0$ is a fixed point of the iterated
map $f^n$ if $x_0$ is a periodic point of period $n.$ 

Simplest dynamical laws are given by monomial functions $f_s(x)= x^s,s=2,3,...$
(each branch of the $p$-adic tree is multiplied by itself $s$ times producing a new branch).

Our basic idea is associate with the genetic code some polynomial
$$
f_{\rm{genetic}}(x)= a_0 + a_1 x +... + a_n x^n,  x  \in {\bf Z}_m,
$$
where depending on the choice of the coding system $m=4, 2.$  

Such a polynomial encodes amino acids in the following way. The set of codons (which are 
considered as 2-adic numbers) is split by this polynomial into groups of cycles. Each cycle 
encodes one amino acid, so:

{\it Amino acids are coded by cycles of this polynomial.} 

Our model cannot explain the origin of such a coding polynomial. Its origin 
can be a consequence of biological evolution or just purely information features of the genetic 
system. Since we do not know the (e.g., biological)  background  inducing a coding polynomial
$f_{\rm{genetic}}(x).$ We are not able to choose it in the unique way. 
In this note we propose one of possible solutions of the problem of finding 
of a coding polynomial.
 
We shall use the well known Mahler's polynomials. To proceed in this way, we choose the 2-adic genetic
coding. The $m=2$ is a prime number and the system of 2-adic integers ${\bf Z}_2$ can be extended 
to the field of 2-adic numbers ${\bf Q}_2.$ We recall that in a number field one can use 
all arithmetic operations: addition, subtraction, multiplication and division. We need these 
operations to define a Mahler's polynomial (the main problem is division). It would be a map 
$f_{\rm{genetic}}: {\bf Z}_2 \to {\bf Q}_2$ having the structure of cycles corresponding to the 
genetic code of amino acids.

Let we know values of some function $f: {\bf Z}_2 \to {\bf Q}_2$ 
in points $j=0,1,...,n.$ Then its $n$th Mahler coefficient is defined by 
$$
a_n= \sum_{j=0}^n (-1)^{n-j} \binom{n}{j} f(j).
$$
The corresponding Mahler's polynomial has the form:
$$
F_n(x)= \sum_{k=0}^n a_k \binom{x}{k},
$$
where the binomial polynomial
$$
\binom{x}{k}= \frac{x(x-1) (x-2)...(x-k+1)}{k!}
$$
The crucial is that 
$$
f(j)= F_n(j), j=0,1,...,n.
$$

Coming back to the genetic code, we see that there are 64 different points-codons. 
Thus we need a Mahler polynomial of degree 63 such that
 
\[\rm{Met}: f(010011)= 010011;  \rm{Trp}:f(001111)= 001111;\]
\[\rm{Asn:} f(010100)= 010110, f(010110)= 010100;\]
\[\rm{Asp:} f(110100)=110110, f(110110)=110100,...,\]

\[\rm{Ser:} f(001000)= 001010, f(001010)= 001001, f(001001)= 000111,\] 
\[f(000111)= 011100, 
f(011100)= 011110, f(011110)= 001000.\]

\section{Representation of gene code through dynamics of fuzzy cycles}

By using the 2-adic coding we can represent each codon with a 2-adic ball of the radius
$r=1/64$ with center in the corresponding 2-adic word.
For example, $010011 \to U_{1/64}(010011).$ This is the set of all 
2-adic sequences such that the first sixth digits coincides with the codon word
$010011.$ Thus the amino acid Met can be represented by the ball $U_{1/64}( 010011)$
and  Trp by  $U_{1/64}(001111).$ But Asn by the union of two balls:
$U_{1/64}(010100) \cup U_{1/64}(010110)$ and, e.g., Ser by the union of sixth
balls  $$
U_{1/64}(001000)\cup U_{1/64}(001010)\cup U_{1/64}(001001)
$$
$$ \cup U_{1/64}(000111) 
\cup U_{1/64}(011100) \cup U_{1/64}(011110).
$$
We remark that in Dragovich, 2006, there was considered a 5-adic model to explain the 
origin of the gene code. In this model 5-adic balls were  used to classify codons.

In Khrennikov, 1997, there were also considered {\it fuzzy cycles,}
cycles of balls,
$$
U_{r_1}(a_1)\to U_{r_2}(a_2) \to\cdots
\to U_{r_k}(a_k) \to U_{r_1}(a_1).
$$ 
We can easily define the notion of attractor fuzzy cycle and Siegel fuzzy cycle.
The basin of attraction of a fuzzy cycle is a set of all points 
which are attracted by such a cycle.

As we have seen in Dubischar et al, 1999, and Khrennikov and Nilsson, 2004, 
consideration of fuzzy cycles is a more natural, since they are stable 
with respect to noise (ordinary cycles can be easily disturbed by noisy 
perturbations).  Now we consider a model in that the ``genetic polynomial'' 
$f_{\rm{genetic}}(x)$ encodes amino acids in the following way:
{\it Amino acids are coded by fuzzy cycles of this polynomial.} However, 
at the moment we do not have mathematical examples of simple polynomials 
having the structure of fuzzy cycles corresponding to the genetic code. We 
shall continue the study of this problem in one of our following papers.

{\bf Conclusion.} {\it We have seen that the genetic code has a natural 
4-adic (or 2-adic) structure. Gene expression could be coupled to a 
dynamical system in the  genetic information space.}

\bigskip

{\bf REFERENCES}
 
Avetisov, V.A., Bikulov, A. H., Kozyrev, S. V., 1999a, 
Description of Logarithmic Relaxation by a Hierarchical Random Walk Model.
Doklady. Mathematics 60 (2), 271-274
 
Avetisov, V. A.,  Bikulov, A. H.,  Kozyrev, S. V., 1999b,
Application of p-adic analysis to models of spontaneous 
breaking of the replica symmetry, arXiv.org:cond-mat/9904360

Avetisov, V. A.,   Bikulov, A. H.,  Kozyrev,   S. V.,  and 
Osipov, V. A., 2002a,  p-adic models of ultrametric diffusion constrained 
by hierarchical energy landscapes. J. Phys. A: Math. Gen. 35,  177-189.

Avetisov, V. A., Bikulov, A. H.,  Osipov, V. A., 2002b,
p-Adic description of characteristic relaxation in complex systems,
arXiv.org:cond-mat/0210447

Beltrametti, E., and Cassinelli, G., 1972,  Quantum
mechanics and $p$-adic numbers. Found. of Physics 2, 1--7.

Dragovich, B.,  and Dragovich, A.,  2006, A p-Adic model of DNA sequence and genetic code, 
q-bio.GN/0607018.

Dubischar D., Gundlach V.M., Steinkamp O.,
Khrennikov A. Yu., 1999, A $p$-adic model for the process of thinking
disturbed by physiological and information noise. J. Theor. Biology,
197, 451-467.

Khrennikov, A. Yu., 1994,  $p$-adic valued distributions in  mathematical physics.
Kluwer Academic Publishers, Dordrech.

Khrennikov, A. Yu. , 1997.  Non-Archimedean analysis: quantum
paradoxes, dynamical systems and biological models. Kluwer,
Dordrecht.

Khrennikov, A. Yu.,  1998a.  Human subconscious as the $p$-adic
dynamical system. J. of Theor. Biology 193, 179-196.

Khrennikov, A. Yu.,  1998b.  $p$-adic dynamical systems: description
of concurrent struggle in biological population with limited growth.
Dokl. Akad. Nauk 361, 752.

Khrennikov, A. Yu., 1999, Description of the operation of the human
subconscious by means of $p$-adic dynamical systems. Dokl. Akad.
Nauk 365, 458-460.

Khrennikov, A. Yu., 2000a,   $p$-adic discrete dynamical systems and
collective behaviour of information states in cognitive models.
Discrete Dynamics in Nature and Society 5, 59-69.

Khrennikov,  A. Yu., 2000b.  Classical and quantum mechanics on
$p$-adic trees of ideas. BioSystems   56, 95-120.

Khrennikov, A.Yu.,  2004a, Information dynamics in cognitive,
psychological, social,  and anomalous phenomena. Kluwer, Dordreht.

Khrennikov, A. Yu., 2004b, Probabilistic pathway representation of
cognitive information. J. Theor. Biology 231, 597-613.

Khrennikov, A. Yu., 2006a,  Ultrametric thinking and Freud's theory of unconscious
mind. In: New Research on Consciousness,  editor Locks J. T., 
Nova Science Publ., Inc, 117-185.

Khrennikov, A. Yu., 2006b, P-adic information space and gene expression. In: Integrative 
approaches to brain complexity, editors Grant S., Heintz N., Noebels J., Wellcome Trust Publ., p.14. 

Khrennikov,  A. Yu.,  Kozyrev, S. V., 2006a,  Replica symmetry breaking
related to a general ultrametric space I: replica matrices and
functionals. Physica A  359, 222-240.

Khrennikov, A. Yu., Kozyrev,  S. V., 2006b,     Replica symmetry breaking
related to a general ultrametric space II: RSB solutions and the
$n\to0$ limit.  Physica A  359, 241-266.

Khrennikov A.Yu., Nilsson M., 2004, $p$-adic deterministic and
random dynamical systems. Kluwer, Dordreht.

Kozyrev, S. V.,  Osipov, V. A.,   Avetisov, V. A., 
2005,  Nondegenerate ultrametric diffusion. Journal of Mathematical Physics
46, 63302. 

Nature Collection, 2006, Human Genom, 1 June 2006.

Parisi,   G., Sourlas, N., 2000,   $p$-adic numbers and replica symmetry breaking.
Europ. Phys. J. 14B, 535-542.

Pitk\"anen, M., 1998. TGD inspired theory of consiousness with
applications to biosystems. Electronic book:
http://www.physics.helsinki.fi/~matpitka/cbookI.html

Voronkov, G.S., 2002a. Information and brain: viewpoint of
neurophysiolog. Neurocomputers: development and applications N1-2,
79-88.

Voronkov, G.S.,  2002b.  Why is the perceived visual world
non-mirror? Int. J. of Psychophysiology 34, 124-131.

Wikipedia, 2006, Genetic code, http://en.wikipedia.org.

Vladimirov, V. S.,  Volovich, I. V., and Zelenov, E. I., 1994,  $p$-adic
Analysis and Mathematical Physics. World Scientific Publ., Singapore.

Volovich, I. V., 1987,  $p$-adic string. Class. Quant. Gravity 
4, 83--87.

\end{document}